\documentclass[floats,prb,aps,12pt]{revtex4}
\usepackage{graphicx}
\begin{document}

\date{\today}
\vspace{2.7in}

\title{Theory of Bose-Einstein condensation and superfluidity of two-dimensional
polaritons in an in-plane harmonic potential}

\author{Oleg L. Berman$^{1}$, Yurii E. Lozovik$^{2}$ and  David W.
Snoke$^{3}$}

\affiliation{\mbox{$^{1}$Physics Department, New York City College
of Technology of the City University of New York,} \\ 300 Jay
Street, Brooklyn, NY 11201 \\
 \mbox{$^{2}$ Institute of
Spectroscopy, Russian Academy of Sciences,}  \\ 142190 Troitsk,
Moscow Region, Russia \\
\mbox{$^{3}$Department of Physics and Astronomy, University of
Pittsburgh,}  \\ 3941 O'Hara Street, Pittsburgh, PA 15260 USA  }

\date{\today}

\begin{abstract}

Recent experiments have shown that it is possible to create an in-plane harmonic potential trap for a two-dimensional (2D) gas of exciton-polaritons
in a microcavity structure, and evidence has been reported of Bose-Einstein condensation of polaritons accumulated in this type of trap.
We present here the theory of Bose-Einstein condensation (BEC) and superfluidity of the exciton polaritons in a harmonic potential trap.
Along the way, we determine a general method for defining the
superfluid fraction in a 2D trap, in terms of angular momentum representation. We show that in the continuum limit, as the trap becomes
 shallower the superfluid fraction approaches the 2D Kosterlitz-Thouless limit, while the condensate fraction approaches zero, as expected.

\vspace{0.1cm}



\end{abstract}

\maketitle {}


\section{Introduction}
\label{intro}

In the past decade, there has been extensive work on Bose coherent effects of two-dimensional (2D) exciton-polaritons in cavitie. (For general reviews see Refs.
\onlinecite{pssb} and \onlinecite{book}). A microcavity is formed by two mirrors opposite each other, as in a laser cavity, with quantum wells embedded in the cavity at
the antinodes of the confined optical mode. The resonant exciton-photon coupling leads to two polariton branches in the spectrum. The lower polariton (LP) branch has a
minimum at $k = 0$ with a very small effective mass, in the range $10^{-5}-10^{-4}$ of the vacuum electron mass, depending on the details of the structure. These
quasiparticles act as a weakly interacting gas of bosons in two dimensions. Since the thermal deBroglie wavelength in two dimensions varies inversely with mass,  the extremely light mass of these bosonic particles means that the critical
 temperature for superfluidity can in principle be 100 K or above for experimentally achievable number densities.

In a translationally invariant two-dimensional system, without a trap,
superfluidity occurs via a Kosterlitz-Thouless superfluid (KTS) transition.  Experiments on untrapped systems \cite{yama,pnas,dang1} have shown promising indications of
 the onset of spontaneous coherence effects. This can be viewed as a type of Bose-Einstein condensation (BEC), with coherence length on the order of the size of the cloud of particles, what is sometimes called a ``quasicondensate.''\cite{kavsst}  It is possible, however, to have a true Bose-Einstein condensation (BEC)  quantum phase transition in two dimensions, if there is a confining potential.\cite{Bagnato,Nozieres} Recently, an experimental method has been demonstrated for creating
such a confining potential trap in a 2D exciton-polariton system, in which the exciton energy is shifted using a stress-induced band-gap shift \cite{Balili}, and evidence for Bose-Einstein condensation of polaritons has been observed in this system \cite{science}. In these experiments, the trap is macroscopic, about 30 microns across compared to a typical interparticle distance of 0.3 microns, and the spring constant is low enough that the spacing between the quantized states $\hbar\omega_0$ in the harmonic potential is small compared to $k_BT$, so that the states may be treated as a continuum.  The diffusion length of the polaritons is comparable to the trap size, so that we may consider them to be in equilibrium spatially.

The properties of polaritons have been studied in several theoretical works. The theory of polariton dynamics due to polariton-polariton interaction has been developed in
Refs.~\onlinecite{Ciuti-exex,Tassone,Ciuti,Porras}. The crossover between lasing and polariton coherence has been studied in Refs.~\onlinecite{Eastham} and
\onlinecite{keeling}. Polariton superfluidity has been predicted\cite{Carusotto} as well as spontaneous linear polarization of the light emission.\cite{kav-lin} In these
previous studies, the coherent polaritonic phases were analyzed in the 2D infinite system.  

 In this paper we present the theory of the trapped polariton condensate.
The paper is organized in the following way.  In Sec.~\ref{pol_ch} the effective Hamiltonian of of microcavity polaritons in trapping potential is derived. In
Sec.~\ref{bec_ch} the number of polaritons in Bose-Einstein condensate (BEC) as a function of temperatures is calculated. The superfluid fraction as a function of
temperature is also obtained. Finally, in Sec.~\ref{discussion} we present our conclusions.

\section{The effective Hamiltonian of of microcavity polaritons in trapping potential}
\label{pol_ch}

The polaritons are linear superpositions of excitons and photons. The Hamiltonian of polaritons is given by $\hat{H}_{tot} = \hat{H}_{exc} + \hat{H}_{ph} + \hat{H}_{exc-ph}$, where
$\hat{H}_{exc}$ is an excitonic Hamiltonian, $\hat{H}_{ph}$ is a photonic Hamiltonian, $\hat{H}_{exc-ph}$ is a Hamiltonian of exciton-photon interaction.  Analogous to the
case of Bose atoms in a trap,\cite{Pitaevskii,Mullin}  in the case of a slowly varying external potential, we can make the quasiclassical approximation, assuming that the
effective exciton mass is not a function of $r$. This quasiclassical approach is valid only if the characteristic $Pr \gg \hbar$, where $P$ is the momentum and $r$ is the radial coordinate in the trap. This is the case in the recent experiments.\cite{science}

The Hamiltonian of $2D$  excitons in the infinite
homogeneous system is given by
\begin{eqnarray}
\label{Ham_exc} \hat H_{exc} = \sum_{{\bf P}}^{}\varepsilon_{ex}(P) \hat{b}_{{\bf P}}^{\dagger}\hat{b}_{{\bf P}}^{} + \frac{1}{2A}\sum_{{\bf P},{\bf P}',{\bf q}}
U_{\bf{q}}
 \hat{b}_{{\bf P} +{\bf q}}^{\dagger}\hat{b}_{{\bf P}'- {\bf q}}^{\dagger}\hat{b}_{{\bf P}}\hat{b}_{{\bf P}'},
\end{eqnarray}
where $\hat{b}_{{\bf P}}^{\dagger}$ and $\hat{b}_{{\bf P}}$ are excitonic creation and
annihilation operators obeying to Bose commutation relations. In the first term, $\varepsilon_{ex}(P) = E_{band} - E_{binding} + \varepsilon
_{0}(P)$ is the energy dispersion of a single exciton in a quantum well, where $E_{band}$ is the band gap energy energy, $E_{binding} = Ry_{2}^{*} =
\mu_{e-h}e^{4}/(\hbar^{2}\epsilon)$ is the binding energy of a  2D exciton ($\mu_{e-h} = m_{e}m_{h}/(m_{e} + m_{h})$ is the reduced excitonic mass, $\epsilon$ is the dielectric constant, and $e$ is the charge of an electron), and  $\varepsilon _{0}(p) = P^2/(2M)$, where $M = m_{e} + m_{h}$ is the
mass of an exciton. In the second, interaction term, $A$ is the macroscopic quantization area and $U_{\bf{q}}$ is the Fourier transform of the exciton-exciton pair repulsion potential. As discussed in Refs.~\onlinecite{Ciuti-exex} and \onlinecite{Laikht},
in the low-density limit, the excitons can be treated as pure bosons, with an interaction potential that includes effects of the underlying fermion nature of the
electrons and holes. For small wave vectors ($q \ll a_{2D}^{-1}$, where $a_{2D} = \hbar ^{2}\epsilon/(2\mu_{e-h}e^{2})$ is the effective 2D Bohr radius of excitons) the
pair exciton-exciton repulsion can be approximated as a contact potential $U_{\bf{q}} \simeq U_{0} \equiv U = 6e^{2}a_{2D}/\epsilon$. This approximation for the exciton-exciton
repulsion is applicable, because resonantly excited excitons have very small wave vectors.\cite{Ciuti} Another reason for this approximation is that the exciton gas is
assumed to be very dilute and the average distance between excitons $r_{s} \sim (\pi n)^{-1/2} \gg a_{2D}$, which implies the characteristic momentum $q \sim
r_{s}^{-1} \ll a_{2D}^{-1}$. A much smaller contribution to the exciton-exciton interaction is also given by band-filling saturation effects,\cite{Rochat} which are neglected
here.

\begin{figure}[t]
   \centering
   \includegraphics[width=3in]{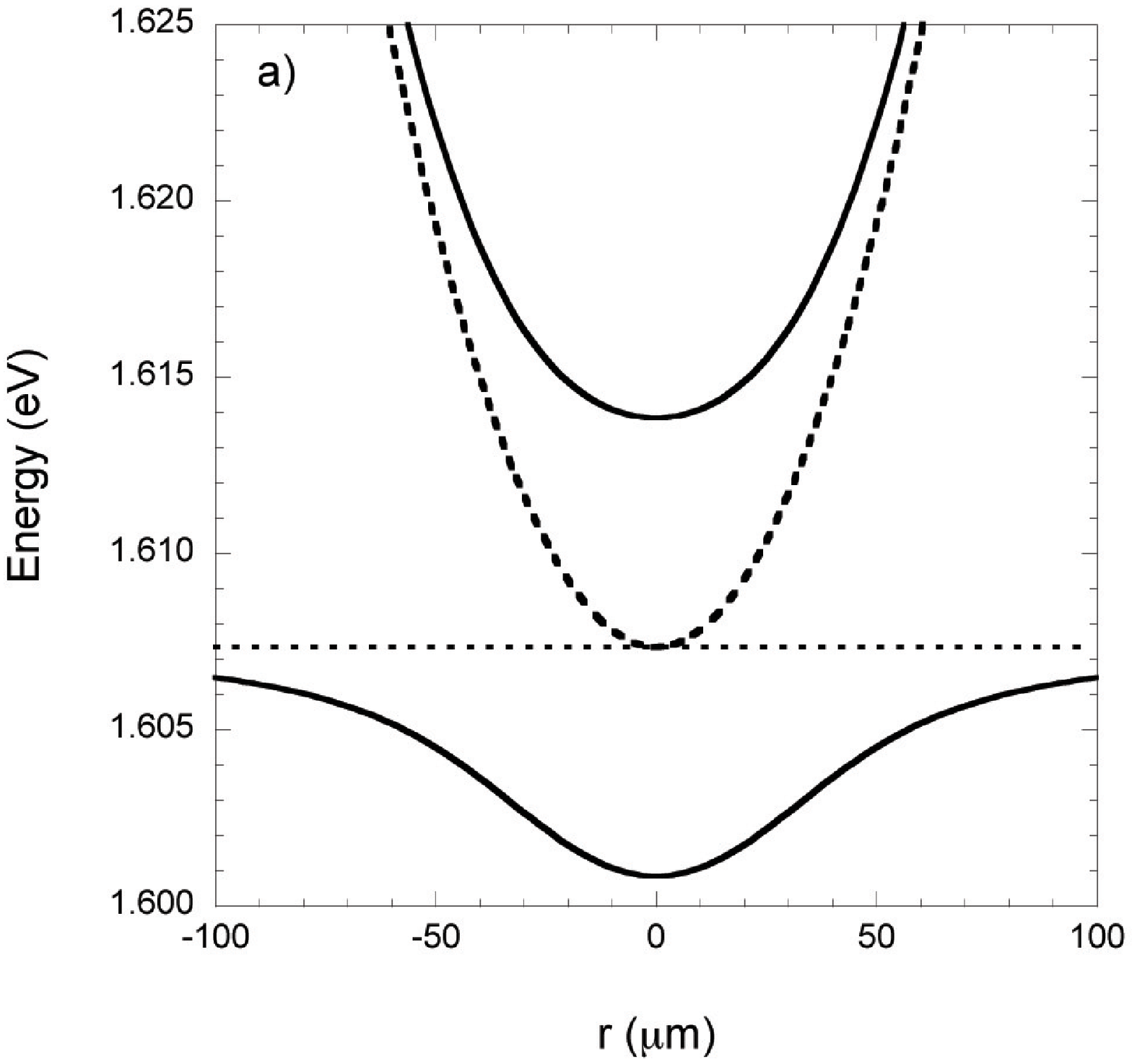}
     \includegraphics[width=3in]{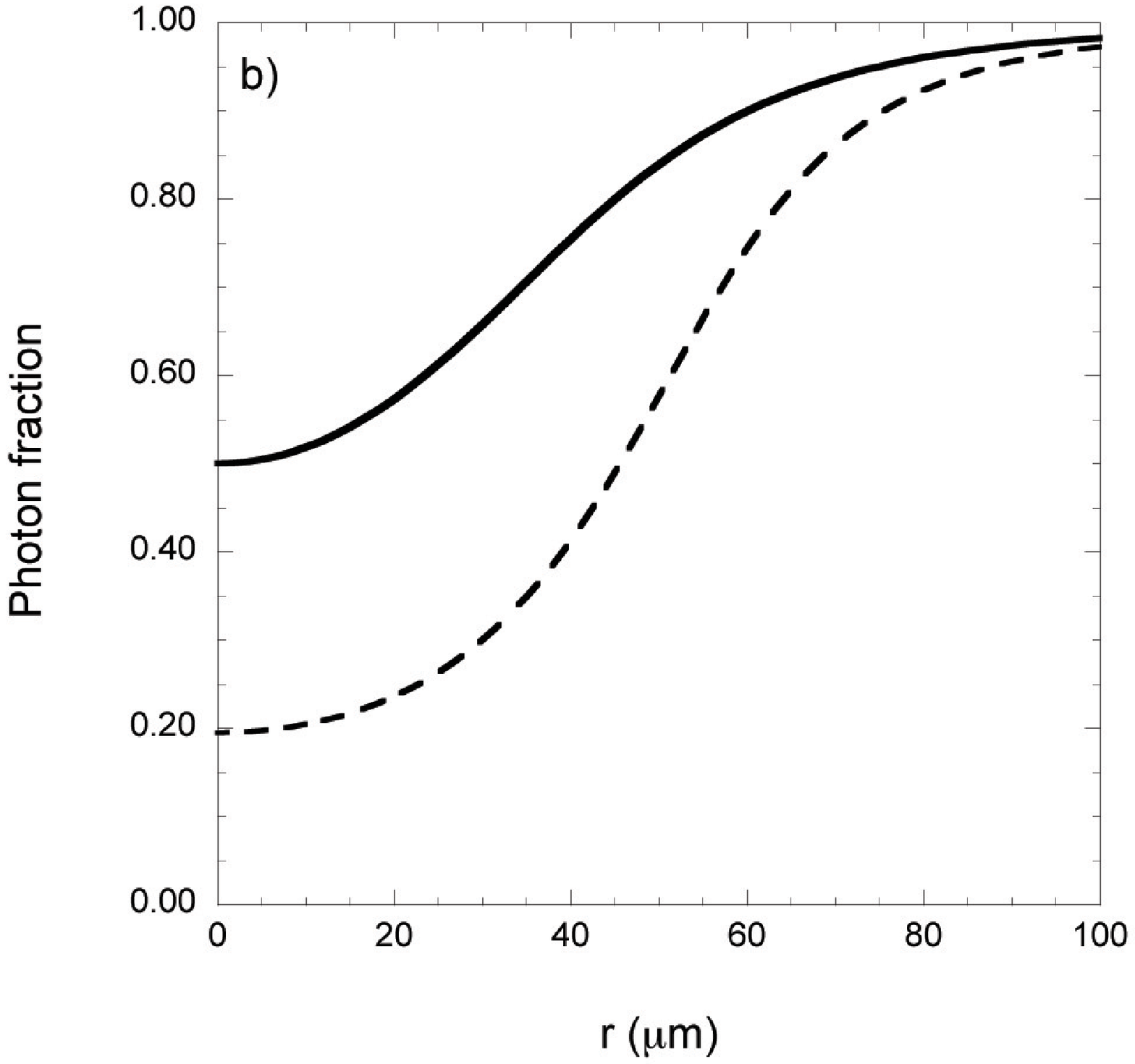}
   \caption{a) Dashed line: the energy of the bare exciton vs. $r$, for the parameters of the GaAs/AlGaAs structure described in the text. Dotted line: the energy of the
   cavity photon mode. Solid lines: the upper and lower polariton energy which arises from the mixing of the photon and exciton modes. b) The photon fraction as a function
   of $r$, for two cases: solid line: exciton energy resonant with the cavity photon mode at $r=0$; dashed line: exciton energy detuned 10 meV below the cavity photon.}
   \label{fig:example1}
\end{figure}
The spatial dependence of the external field $V(r)$ comes about due to the shifting of the exciton energy with inhomogeneous stress;\cite{Snoke-apl}  the photon states in
the cavity are assumed to be unaffected by stress.  In this case the band energy $E_{band}$ is replaced by $E_{band}(r) = E_{band}(0)+ V(r)$. Near the minimum of the exciton energy, $V(r)$ can be approximated as $\frac{1}{2} \gamma  r^{2}$.

The Hamiltonian of non-interacting photons in a semiconductor microcavity is given by\cite{Pau}:
\begin{eqnarray}
\label{Ham_ph} \hat H_{ph} = \sum_{{\bf P}}  \varepsilon _{ph}(P) \hat{a}_{{\bf P}}^{\dagger}\hat{a}_{{\bf P}}^{} ,
\end{eqnarray}
where $\hat{a}_{{\bf P}}^{\dagger}$ and $\hat{a}_{{\bf P}}$ are photonic creation and
annihilation Bose operators, and $\varepsilon _{ph}(P) = (c/n)\sqrt{P^{2} + \hbar^{2}\pi^{2}L_{C}^{-2}}$ is the cavity photon spectrum ($c$ is the speed of light in
vacuum, $L_{C}$ is the length of the cavity, and $n = \sqrt{\epsilon}$ is the effective refractive index).

The Hamiltonian of harmonic exciton-photon coupling has the form:\cite{Ciuti}
\begin{eqnarray}
\label{Ham_exph} \hat{H}_{exc-ph} = {\hbar \Omega_{R}}\sum_{{\bf P}}
 \hat{a}_{{\bf P}}^{\dagger}\hat{b}_{{\bf P}}^{} + h.c. ,
\end{eqnarray}
where the exciton-photon coupling energy represented by the Rabi constant $\hbar \Omega_{R}$ depends on the overlap between the exciton and photon wavefunction and the semiconductor dipole moment.\cite{Savona_review} We neglect anharmonic terms for the exciton-photon coupling.

The linear part of the total Hamiltonian $\hat{H}_{tot}$ (without the second term on the right-hand side of Eq.~(\ref{Ham_exc})) can be diagonalized by applying unitary
transformations and has the form:\cite{Ciuti}
\begin{eqnarray}
\label{h0} \hat{H}_{0} = \sum_{\mathbf{P}}\varepsilon_{LP}(P)\hat{p}_{\mathbf{P}}^{\dagger}\hat{p}_{\mathbf{P}}
+\sum_{\mathbf{P}}\varepsilon_{UP}(P)\hat{u}_{\mathbf{P}}^{\dagger}\hat{u}_{\mathbf{P}},
\end{eqnarray}
where $\hat{p}_{\mathbf{P}}^{\dagger}$ and $\hat{u}_{\mathbf{P}}^{\dagger}$ are the Bose
creation and operators for the lower and upper polaritons, respectively; the energy spectra of the low/upper polaritons are
\begin{eqnarray}
\label{eps0} \varepsilon_{LP/UP}(P) = \frac{\varepsilon _{ph}(P) +\varepsilon _{ex}(P)}{2} \mp \frac{1}{2}\sqrt{(\varepsilon_{ph}(P) - \varepsilon _{ex}(P))^{2} +
4|\hbar\Omega_{R}|^{2}},
\end{eqnarray}
which implies a splitting between the upper and lower states at
$P=0$ of $2\Omega_R$, known as the Rabi splitting.  For the GaAs
cavities used in Ref. \onlinecite{science}, this splitting was
approximately 14 meV. The upper and lower polariton energies from
(\ref{eps0}) are plotted in Figure 1(a) as a function of $r$ at
momentum $P=0$ for a value of $\gamma$ chosen to give
 a fit to the experimentally measured curvature of the lower polariton branch.  The fit implies $\gamma = 960$ eV/cm$^2$ for the bare excitons.

The excitonic and photonic operators are defined as\cite{Ciuti}
\begin{eqnarray}
\label{bog_tr} \hat{b}_{\mathbf{P}} = X_{P}\hat{p}_{\mathbf{P}} - C_{P}\hat{u}_{\mathbf{P}}, \hspace{3cm}
 \hat{a}_{\mathbf{P}} = C_{P}\hat{p}_{\mathbf{P}} +
X_{P}\hat{u}_{\mathbf{P}},
\end{eqnarray}
 where $\hat{p}_{\mathbf{P}}$ and $\hat{u}_{\mathbf{P}}$ are lower and upper polariton Bose
operators, respectively, and $X_{P}$ and $C_{P}$ are\cite{Ciuti}
\begin{eqnarray}
\label{bog} X_{P} = \frac{1}{\sqrt{1 + \left(\frac{\hbar\Omega_{R}}{\varepsilon_{LP}(P) -
 \varepsilon _{ph}(P)}\right)^{2}}} , \hspace{3cm}
C_{P} = - \frac{1}{\sqrt{1 + \left(\frac{\varepsilon_{LP}(P) -
 \varepsilon _{ph}(P)}{\hbar\Omega_{R}}\right)^{2}}} ,
\end{eqnarray}
where $|X_{P}|^{2}$ and $|C_{P}|^{2} = 1 - |X_{P}|^{2}$ represent the exciton and cavity photon
fractions in the lower polariton.\cite{Ciuti} Figure 1(b) shows the photon fraction at zone center, $|C_0|^2$, as a function of $r$. Further from the center, the exciton energy $\varepsilon_{ex}$ becomes detuned from the cavity photon energy, leading the lower polariton to become more photon-like. Because cavity photon lifetime is so much shorter than the intrinsic exciton lifetime ($\sim 2$ ps compared to $\sim 100$ ps), the polariton lifetime is proportional to the photon fraction. This implies that polaritons at higher energy in the trap have shorter lifetime; in other words, there is an evaporative cooling effect.   As shown in Fig.~1(b), this effect can be magnified by tuning the exciton level below the photon level at the center of the trap, so that the polaritons are more excitonic there.

Substituting the polaritonic representation of the excitonic and photonic operators (\ref{bog_tr}) into the total Hamiltonian
$\hat{H}_{tot}$, the Hamiltonian of lower polaritons is obtained:\cite{Ciuti}
\begin{eqnarray}
\label{Ham_tot_p} \hat H_{tot} = \sum_{\mathbf{P}}\varepsilon_{LP}(P)\hat{p}_{\mathbf{P}}^{\dagger}\hat{p}_{\mathbf{P}} + \frac{1}{2A}\sum_{{\bf P},{\bf P}',{\bf q}}
U_{\mathbf{P},\mathbf{P}',\mathbf{q}}
 \hat{p}_{{\bf P} +{\bf q}}^{\dagger}\hat{p}_{{\bf P}'- {\bf q}}^{\dagger}\hat{p}_{{\bf P}}\hat{p}_{{\bf P}'},
\end{eqnarray}
where
\begin{eqnarray}
\label{U_p} U_{\mathbf{P},\mathbf{P}',\mathbf{q}} = \frac{6e^{2}a_{2D}}{\epsilon }X_{{\bf P} +{\bf q}} X_{{\bf P}'}X_{{\bf P}'-{\bf q}}X_{{\bf P}}.
\end{eqnarray}

For the slowly changing confinement potential $V(r) = - (E_{band} - E_{binding}) + (c/n) \hbar\pi L_{C}^{-1} + \frac{1}{2}\gamma
 r^{2}$ ($r$ is the distance between the center of mass of the
   exciton and the center of the trap), the exciton spectrum is
   given in the effective mass approximation as
\begin{eqnarray}
\label{ex_sp} \varepsilon_{ex}^{(0)}(P) = \varepsilon_{ex}(P) + V(r) = (c/n) \hbar\pi L_{C}^{-1} +  \frac{\gamma}{2}r^{2} + \frac{P^{2}}{2M}.
\end{eqnarray}
This quasiclassical approximation is valid if $P >> \hbar/l$, where
$l = \left(\hbar/(M\omega_{0})\right)^{1/2}$ is the size of the
exciton cloud in ideal exciton gas and $\omega_{0} =
\sqrt{\gamma/M}$.

At small momenta $\alpha \equiv 1/2 (M^{-1} +
(c/n)L_{C}/\hbar\pi)P^{2}/|\hbar \Omega_{R}| \ll 1$ and weak
confinement potential $\beta \equiv \gamma r^{2}/|\hbar \Omega_{R}|
\ll 1$, the single-particle lower polariton spectrum obtained by
substitution of Eq.~(\ref{ex_sp}) into Eq.~(\ref{eps0}), in linear
order with respect to the small parameters $\alpha$ and $\beta$, is
\begin{eqnarray}
\label{eps00} \varepsilon_{0}(P) \approx (c/n) \hbar \pi L_{C}^{-1}  - |\hbar \Omega_{R}| +  \frac{\gamma}{4} r ^{2} + \frac{1}{4}   (M^{-1} + (c/n)L_{C}/\hbar\pi)P^{2} .
\end{eqnarray}
By substituting Eq.~(\ref{ex_sp}) into Eq.~(\ref{bog}) we obtain $X_{\mathbf{P}} \approx
1/\sqrt{2}$. The condition of the validity of the quasiclassical approach in Eq.~(\ref{Ham_exc}), $Pr\gg\hbar$, is also applied here.

If we measure energy relative to the $P=0$ lower polariton energy $(c/n) \hbar \pi L_{C}^{-1}  - |\hbar \Omega_{R}|$, the resulting effective Hamiltonian for polaritons in the parabolic trap in $\mathbf{P}$
space in the effective mass approximation has the form:
\begin{eqnarray}
\label{Ham_eff} \hat H_{\rm eff}  = \sum_{\mathbf{P}}\left(\frac{P^{2}}{2M_{\rm eff}} + V_{\rm eff}(r) \right)\hat{p}_{\mathbf{P}}^{\dagger}\hat{p}_{\mathbf{P}} +
\frac{U_{\rm eff}^{(0)}}{2A}\sum_{{\bf P},{\bf P}',{\bf q}}
 \hat{p}_{{\bf P} +{\bf q}}^{\dagger}\hat{p}_{{\bf P}'- {\bf q}}^{\dagger}\hat{p}_{{\bf P}}\hat{p}_{{\bf P}'},
\end{eqnarray}
 where the sum over $\mathbf{P}$ and $\mathbf{P}'$ is carried out only over $P >> \hbar/l$
  (as only in this case the quasiclassical approximation used in Eq.~(\ref{ex_sp}) is valid), and
 the effective mass of a polariton is
given by
\begin{eqnarray}
\label{Meff} M_{\rm eff}^{-1} = \frac{1}{2}   (M^{-1} + (c/n)L_{C}/\hbar\pi) ;
\end{eqnarray}
the effective external potential $V_{\rm eff}(r) = \frac{1}{2}V(r)$
(i.e., $\gamma_{\rm eff} =  \gamma/2$), and the effective
polariton-polariton pair repulsion potential is given by the
hard-core contact potential $U_{\rm eff}(\mathbf{r} - \mathbf{r}') =
U_{\rm eff}^{(0)} \delta(\mathbf{r} - \mathbf{r}') =
\frac{1}{4}U_0\delta(\mathbf{r} - \mathbf{r}')$. Using the
experimental parameters for GaAs/AlGaAs structure used in
Ref.~\onlinecite{science} ($E_{ph} = 1.60735$ eV) we obtain $M_{\rm
eff} = 7.8\times10^{-5} m_0$, where $m_0$ is the vacuum electron
mass, in good agreement with the value of  $7\times10^{-5} m_0$
obtained from direct measurement of the effective polariton mass
using angle-resolved photon detection, reported in
Ref.~\onlinecite{science}.

\section{Bose-Einstein condensation and superfluidity of microcavity polaritons}
\label{bec_ch}

In the real space the effective Hamiltonian for trapped polaritons will look exactly like the Hamiltonian of weakly-interacting dilute 2D Bose gas in a confinement:
\begin{eqnarray}
\label{Ham_eff_real} \hat H_{\rm eff}  = \int d\mathbf{r} \hat{\psi}^{\dagger}(\mathbf{r})\left( - \frac{\hbar^2\nabla^2}{2M_{\rm eff}} + V_{\rm eff}(r) \right)\hat{\psi}(\mathbf{r}) +
\frac{U_{\rm eff}^{(0)}}{2} \int d\mathbf{r} \hat{\psi}^{\dagger}(\mathbf{r})\hat{\psi}^{\dagger}(\mathbf{r})\hat{\psi}(\mathbf{r})\hat{\psi}(\mathbf{r}),
\end{eqnarray}
where $\hat{\psi}^{\dagger}(\mathbf{r})$ and $\hat{\psi}(\mathbf{r})$ are real space Bose field
operators of creation and annihilation of polaritons, correspondingly.

Although Bose-Einstein condensation (BEC) cannot happen in a 2D homogeneous ideal gas at non-zero temperature, as discussed in Ref.~\onlinecite{Bagnato}, in a harmonic trap BEC can occur in two dimensions below a
critical temperature $T_{c}^{0}$ given by $k_{B} T_{c}^{0}= \hbar \pi ^{-1}\sqrt{6 \gamma_{\rm eff} N/M_{\rm eff}}$, where $k_{B}$ is the
 Boltzmann constant, and $N$ is the total number of polaritons in a trap.
 This expression for the temperature of BEC is valid if we neglect
 the polariton-polariton repulsion, i.e., if we assume $U_{\rm eff}^{(0)} = 0$.

Neglecting the anomalous averages $\left\langle \hat{\psi} \hat{\psi}\right\rangle$  and $\left\langle \hat{\psi}^{\dagger} \hat{\psi}^{\dagger}\right\rangle$ via the Popov
approximation\cite{Giorgini}, implying the system to be very dilute, that is $na_{2D}^{2} \ll 1$, where $n = N/(\pi R^{2})$ is the total density of polaritons, and $R$ is the 2D radius of
the trap, the self-consistent equation for the non-condensate density $n'(r)$ at temperatures $k_{B}T \gg \hbar \sqrt{\gamma_{\rm eff}/M_{\rm eff}}$  can be written as\cite{Mullin}
\begin{eqnarray}
  n'(r) = - \frac{M_{\rm eff} k_{B}T}{2\pi\hbar^{2} }\log\left(1 \nonumber - \exp\left[ -\frac{1}{k_{B}T} \sqrt{ \left( \frac{1}{2} \gamma_{\rm eff} r^2 + 2 U_{\rm eff}^{ (0) } n -
\mu \right)^2 -  | U_{\rm eff}^{ (0) } |^2 n_0^2 }  \right] \right),\nonumber \\ \label{n_ncon}
\end{eqnarray}
where $n_{0}$ is the total density of condensate, and $\mu = 2U_{\rm
eff}^{(0)}n - U_{\rm eff}^{(0)}n_{0}$ is the chemical potential of
the system in the Popov approximation.\cite{Griffin}   For the
experimental parameters of interest,  the size of a trap is $R \sim
30$~$\mu$m, the effective 2D Bohr radius of an exciton is $a_{2D} =
130$ \AA, and $\gamma \sim 10^{3}$~eV/cm$^{2}$, which implies the
above equation is valid for $T \gg 1$~K, which is true in all of
these experiments. Figure 2 plots the spatial profile of the
condensate for the experimental parameters of the trap. Note that
only  the states  $P \gg \hbar/l$ can be treated as quasiclassical.
Since the the characteristic momenta for Bose condensate of weakly
interacting particles is $P = \left( 2 M_{\rm eff} U_{\rm eff}^{ (0)
}n \right)^{1/2}$ satisfy to the condition $\hbar/l \ll P = \left( 2
M_{\rm eff} U_{\rm eff}^{ (0) }n \right)^{1/2}$, we can apply the
quasiclassical approximation.

\begin{figure}[t]
   \centering
   \includegraphics[width=3.5in]{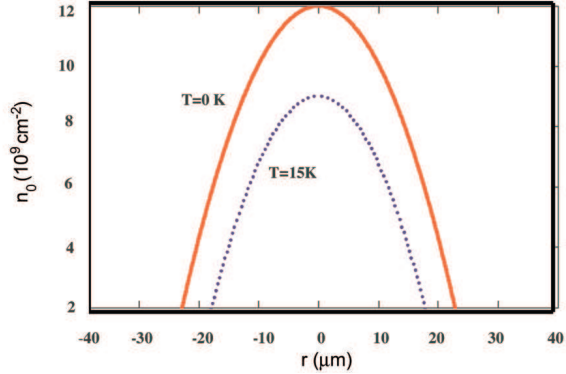}
   \caption{The condensate profile
   $n_0(r)$ in the trap, for $\gamma = 960$ eV/cm$^2$ and total density at the center $n(0) =  1.2\times10^9$ cm$^{-2}$. Solid line: $T = 0$,
   using the Thomas-Fermi/Bogoliubov
   approximation. Dotted line: $T = 15$ K, using the theory presented in the text.
}
\end{figure}

\begin{figure}[htbp] 
   \centering
  \includegraphics[width=3.5in]{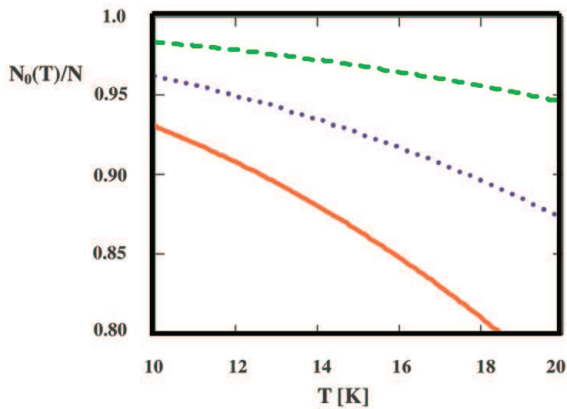}
   \caption{Condensate fraction $N_0/N$ as a function of temperature, for three trap spring constants: solid line: 760 eV/cm$^2$; dotted line: 860 eV/cm$^2$; dashed line:
   960 eV/cm$^2$.}
   \label{fig:example}
\end{figure}

The total number of polaritons in the condensate is given by $N_{0} = N - N'$, where $N' = 2\pi \int_{0}^{R} n'(r) r dr$ is the total number of non-condensate
particles. Assuming $n_{0} = n - n'(r)$, and solving the self-consistent equation (\ref{n_ncon}) with respect to the non-condensate density $n'(r)$, we obtain the
dependence of the fraction $N_{0}(T)/N$ of the total number of condensate particles  on the temperature $T$. This is plotted in Fig.~\ref{fig:example} for the experimental conditions.

In the thermodynamic limit $N \rightarrow \infty$, the Thomas-Fermi approximation that the kinetic energy of the system can be neglected has been proved to be valid for BEC in
a harmonic trap.\cite{Baym} For small quasimomenta $P \ll \sqrt{2M_{\rm eff}U_{\rm eff}^{(0)}n_{0}}$ and small temperatures, the energy spectrum of the quasiparticles $\varepsilon (P)$
is given by\cite{Mullin} $\varepsilon (P,r) \approx c_{s}(r)P$, where $c_{s}(r)$ is the sound velocity in the Popov approximation\cite{Griffin} ($c_{s}(r) =
\sqrt{U_{\rm eff}^{(0)}n_{0}(r,T)/M_{\rm eff}}$).

\begin{figure}[t] 
   \centering
  \includegraphics[width=3.5in]{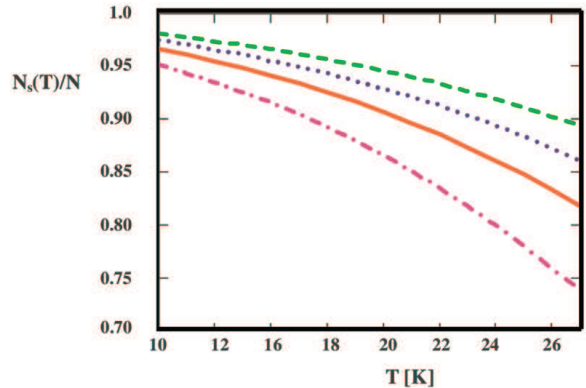}
   \caption{Superfluid fraction $N_s/N$ as a function of temperature for the same
three trap spring constants as Fig.~3. Dashed-dotted line: the superfluid fraction in the limit $\gamma \rightarrow 0$, namely, the translationally invariant 2D case. }
   \label{fig:example4}
\end{figure}

Since the spectrum of the quasiparticles is a linear sound spectrum satisfying the Landau criterium of superfluidity,\cite{Abrikosov} superfluidity of the
polaritons can occur in the trap. Therefore at small temperatures there are two components in the trapped gas of polaritons: the normal component and the superfluid component. We define
the total number of particles in the superfluid component s $N_{s} \equiv N - N_{n}$, where $N_{n}$ is a total number of particles in the normal component. We define $N_{n}$
analogously to the procedure applied for definition of the density of the normal component in the infinite system $n_{n}$,\cite{Abrikosov} using the isotropy of the
trapped polaritonic gas instead of the translational symmetry for an infinite system. We imagine that  a ``gas of quasiparticles'' rotates in the liquid in the plane
perpendicular to the axis of the trap with some small macroscopic angular velocity $\mathbf{\nu}$. In this case, the distribution function of a gas of quasiparticles can
be obtained from the distribution function of a gas at rest by substituting for the energy spectrum of the quasiparticles $\varepsilon (P) - \mathbf{L} \mathbf{\nu}$, where
$\mathbf{L} = \mathbf{r} \times \mathbf{P}$ is the angular momentum of the particle. Assuming $Pr/\hbar \gg 1$, we apply the quasiclassical approximation for the angular
momentum: $L \approx Pr$ and $\varepsilon (L,r) = r^{-1}c_{s}(r)L$. The total angular momentum in a trap per unit of volume $\mathbf{L}_{\rm tot}(r)$ is given by
\begin{eqnarray}
\label{L_tot} \mathbf{L}_{\rm tot}(r) = \int\frac{d^{2}L}{(2\pi \hbar r)^{2}} \mathbf{L} n_{B}\left(\varepsilon(r,L) - \mathbf{L} \mathbf{\nu}\right),
\end{eqnarray}
where we assume that at small temperatures the quasiparticles are non-interacting, and they  are
described by the Bose-Einstein distribution function $n_{B}(\varepsilon) = (\exp[\varepsilon/(k_{B}T)] - 1)^{-1}$. For small angular velocities, $n_{B}\left(\varepsilon -
\mathbf{L} \mathbf{\nu}\right)$ can be expanded with respect to $\mathbf{L} \mathbf{\nu}$. Then we get
\begin{eqnarray}
\label{L_tot_1} \mathbf{L}_{tot}(r) = - \int\frac{d^{2}L}{(2\pi \hbar r)^{2}}  \mathbf{L}(\mathbf{L}\mathbf{\nu})\frac{\partial n_{B}(\varepsilon)}{\partial\varepsilon} ,
\end{eqnarray}
Assuming that only quasiparticles contribute to the total angular momentum, we define the
density of the normal component $n_{n}(r)$ by $\mathbf{L}_{tot}(r) = n_{n}(r)L_{0}$, where $L_{0} = M_{\rm eff}r\nu$ is the angular momentum of one quasiparticle. For the total
number of particles in the normal component we obtain
\begin{eqnarray}
\label{N_n}
 N_{n} = 2\pi \int_{0}^{R} n_{n}(r) r dr =  \int_{0}^{R}
\frac{3 \zeta (3) k_{B}^{3} T^3}{\hbar^{2} c_{s}^{4}(n_{0}(r)) M_{\rm eff}} r dr,
\end{eqnarray}
 where $\zeta (z)$ is the Riemann zeta function ($\zeta (3) \simeq 1.202$), and the density of
the condensate $n_{0}(r) = n - n'(r)$ (the density of non-condensate polaritons $n'(r)$ can be obtained from Eq.~(\ref{n_ncon})). The dependence of the fraction
$N_{s}(T)/N$ of the total number of polaritons in the superfluid component $N_{s} (T) = N - N_{n} (T)$ on the temperature $T$ is presented in Fig.~\ref{fig:example4}. The
superfluid fraction depends only weakly on the spring constant $\gamma$, and in the limit $\gamma \rightarrow 0$ approaches the  the superfluid density for a 2D
translationally invariant system.\cite{berm0}

\section{Discussion}
\label{discussion}

In conclusion, at low temperature, the Hamiltonian of 2D exciton polaritons in a slowly varying external parabolic potential corresponds directly to the case of a weakly interacting Bose gas with an effective mass and effective pair interaction in a
harmonic potential trap. The condensate fraction and the superfluid component are decreasing functions of temperature, as expected, and increasing functions of the
curvature of the parabolic potential.  The mixing with the photon states leads to a decreased lifetime for high-energy states, that is, an evaporative cooling effect, but
does not fundamentally prevent condensation.

The results given here are comparable to those of the experiments, but do not correspond exactly. The condensate peak seen in the experiments with a trap has approximately 15 $\mu$m full width at half maximum, while the peak shown in Fig.~2 comparable conditions has width of approximately 30 $\mu$m. The most likely reason is that the mean-field shift due to the repulsive interaction between particles $U_{\rm eff}n$ is known to strongly overestimate the actual energy shift by as much a factor of ten,\cite{zimm} because anticorrelation of the excitons tends to reduce the average interaction potential.

The condensate fraction obtained is a decreasing function of the characteristic potential of the interparticle repulsion, which corresponds to the results obtained in Ref.~[\onlinecite{Holzmann}]. The authors of Ref.~[\onlinecite{Holzmann}] showed that at finite number of bosons $N$, the interparticle repulsion suppresses the temperature of BEC, and in the thermodynamic limit $N \rightarrow \infty$ the interparticle interaction eliminates BEC at finite temperatures. Since we consider the very dilute  gas of a finite number of polaritons with weak repulsion (weakly-nonideal Bose-gas), the increase of the interparticle repulsion results in the increase of the non-condensate fraction (Eq.~(\ref{n_ncon})) at the fixed finite temperature, which agrees with the results of Ref.~[\onlinecite{Holzmann}].

In our calculations, we have assumed thermal equilibrium. Since the polariton lifetime is short, one may question this assumption. The condition for thermal equilibrium,
however, is simply that the time scale for thermalizing collisions be short compared with the particle lifetime. Porras et al. \cite{Porras} have shown than the time scale
for polariton-exciton scattering can be fast enough for a thermalized distribution of polaritons to exist in lowest $k$-states. Although polaritons have very short
lifetime, thermodynamic equilibrium can be achieved in the regime of the strong pump. Polariton- polariton interactions can help overcome of the bottleneck and lead to large occupation numbers of the ground state. However, we cannot rule out that consideration of pump and decay in a steady state may lead to differences from the results presented here. To give an example, the renormalized dispersion of BEC of particles with infinite lifetime
is Bogoliubov-like, while in the steady state of the system with pump and decay it is very different.\cite{Szymanska_Littlewood} This consideration of the influence of the decay on the BEC may be a subject of further studies of a trapped gas.

The spin polarization is important not only for the excitations, but for the condensate itself. In Ref.~[\onlinecite{Bagnato}] the dynamics of the spin of the polariton
BEC was analyzed in details. It was shown that polariton-polariton interactions lead to the polarization dephasing in spatially confined systems. The influence of spin on
the phase transitions of the trapped polaritonic gases are the subject of further research.


{\bf Acknowledgements}. We thank Vincenzo Savona and Peter B. Littlewood for useful and stimulating discussions. This research has been supported by the National Science
Foundation under Grant No. 0404912 and by DARPA under Army Research Office Contract No. W911NF-04-1-0075. Yu.~E.~L. has been supported by the INTAS and RFBR grants.



\newpage


\end{document}